\pdfoutput=1
\documentclass[pra,aps,preprint,showpacs,nopreprintnumbers]{revtex4}

\usepackage{graphicx}% Include figure files

\usepackage{latexsym}
\usepackage{amsmath}
\usepackage{amsfonts}

\usepackage{exscale}
\usepackage{amssymb}
\usepackage{mathrsfs}
\begin{document}
%\newcommand{\mf}[1]{\boldsymbol{#1}}

%\preprint{APS/123-QED}

\title{Attochirp-free\\ High-order Harmonic Generation }% Force line breaks with \\
\author{Markus~C.~Kohler}
\author{Christoph~H.~Keitel}
\author{Karen~Z.~Hatsagortsyan}
\email{k.hatsagortsyan@mpi-hd.mpg.de}

\address{Max-Planck-Institut f\"ur Kernphysik, Saupfercheckweg 1, 69117 Heidelberg, Germany}

\date{\today}% It is always \today, today,

%explain N_F better,maybe show spectrum
%zs is done relativistically
%get rid of physical energy operator

\begin{abstract}
A method is proposed for arbitrarily engineering the high-order harmonic generation phase achieved by shaping a laser pulse and employing xuv light or x rays for ionization. This renders the production of bandwidth-limited attosecond pulses possible while avoiding the use of filters for chirp compensation. By adding the first 8 Fourier components to a sinusoidal field of $10^{16}$W/cm$^2$, the bandwidth-limited emission of 8 as is shown to be possible from a Li$^{2+}$ gas. The scheme is extendable to the zs-scale.
\end{abstract}

\maketitle

\section{Introduction}
High-order harmonic generation  (HHG) is the key technology for attosecond science.
%to explore physics on a subfemtosecond time scale and, therefore, has become widely accessible tool in attosecond science. 
The shortest pulse durations currently achieved are below 100 as \cite{GOULIELMAKIS2008,SANSONE2010,KO2010}. Highly efficient sources are available  having a bandwidth of hundreds of electronvolts \cite{CHEN2010}  being large enough
to produce pulses of only 10 as duration --- in case they can be generated  without chirp in the future. Many properties of HHG radiation emitted from a gas target can be understood by studying a single atom. The three-step model \cite{CORKUM1993} is the simplest model to describe the single-atom dynamics. The process starts when a strong laser field liberates the electron of an atom and subsequently drives it in the continuum.  If ionization happenend at the right time, the electron can be accelerated back towards the parent ion after the field has changed its sign and can recombine along with the emission of an energetic photon. 

One prominent feature of HHG is that the emitted light has an intrinsic chirp, the so-called attochirp \cite{MAIRESSE2003,KAZAMIAS2004}. 
The origin of the attochirp can be understood from the classical electron trajectories in the laser field: trajectories with different energies recollide at different times. 
%(see dashed gray trajectories in Fig. \ref{figcomparison}). 
%Loosely spoken: different energies are emitted at different times. 
Due to the widespread classical recollision times in a usual sinusoidal laser field, the emitted harmonic pulses have a longer duration than their bandwidth limit, that is the minimum pulse duration for a given spectral bandwidth reached when the spectral phase of the harmonic pulse is constant. To compress the emitted pulse, dispersive elements of either chirped multilayer x-ray mirrors \cite{MORLENS2005},  thin metallic films \cite{KIM2004,LOPEZ2005} or gaseous media \cite{KO2010,KIM2007} are employed and even the use of grating compressors is attempted \cite{POLETTO2008}. However, these techniques 
%lack in flexibility, 
suffer from losses, rely on the material properties and the chirp is not well-controlable in these schemes. Additionally, it is required to select radiation from the positively chirped short trajectory  via phase-matching before the compensation element. Other approaches \cite{CHEN2010,DOUMY2009} try to exploit the reduced time window of recollision when increasing the wavelength of the driving laser at a constant bandwidth and field maximum. However, this way, it cannot be taken advantage of the increased cutoff.  Without selection of a certain trajectory, it is possible to partially reduce the attochirp as shown in~\cite{ZHENG2009,ZOU2009} by  adding a weak second-harmonic or subharmonic field. 
%In fact, for chirp compensation in this case, one has to find a way to fit the larger emitted spectral bandwidth into a smaller fraction of the recollision time window.
The attochirp problem is expected to get more significant and demanding when photon energies spanning far into the soft x-ray domain are reached in the future.
%A broader harmonic spectrum requires a material for chirp compensation which has a suitable dispersion in a larger range. Additionally, for even shorter harmonic bursts than available nowadays, the pulse duration is then only a tiny fraction of the recollision time window making dispersion compensation even more important as discussed later in detail. 
%A special case is the employment of long-wavelength driving lasers \cite{CHEN2010}. By increasing the wavelength the attochirp decreased \cite{DOUMY2009} when the extendend bandwidth is not used. However, when one attempts to use the entire bandwidth the longer period results in a larger resulting in a larger time delay between the trajectories with zero energies and the cutoff. 

%A more elegant solution of this problem would be when the generation process of the single atoms could be altered in a way that the attochirp is engineered in a determined manner or could even vanish. 

In this paper, we propose a way to engineer the attochirp in a determined manner by altering the harmonic generation process.
In terms of the wave function, the following scenario is realized: the wave function localized in the binding potential is continuously freed resulting in a large spread in space and momentum. The electronic quantum dynamics is tuned in a way that after a certain time of propagation, the wave function spatially re-compresses at least along the propagation direction of the wave packet. It has its minimum width exactly at the time of recollision but with a large energy bandwidth gained during  propagation in the continuum. 
We show that when the laser pulse is shaped by adding a small number of low-order harmonics and employing soft x rays for ionization, attosecond pulses with arbitrary chirp can be formed including the possibility of attochirp-free HHG and bandwidth-limited attosecond pulses.

The benefit of assisting the HHG process in a strong laser field with a weak high-frequency field has been demonstrated  mainly for the purpose of enhancing the single-atom yield~\cite{xuv-ass-HHG,SCHAFER2004,TAKAHASHI2007,FIGUEIRA2007,GAARDE2005}, improving phase matching~\cite{GAARDE2005} or suppressing the relativistic drift~\cite{KLAIBER-OL,ufo}. On the other hand, femtosecond pulse shaping has been used to shape the HHG spectrum~\cite{PFEIFER2005}, to increase the HHG cutoff~\cite{CHIPPERFIELD2009} or for relativistic HHG~\cite{klaiber_tailored,klaiber_tailored2}. Here, we employ x rays to ionize the electron with non-zero velocity and combine it with femtosecond pulse shaping to control the spectral phase of the harmonic spectrum.

\section{Classical Analysis}
\begin{figure}%[h]
\begin{center}
\includegraphics[width=0.6\textwidth]{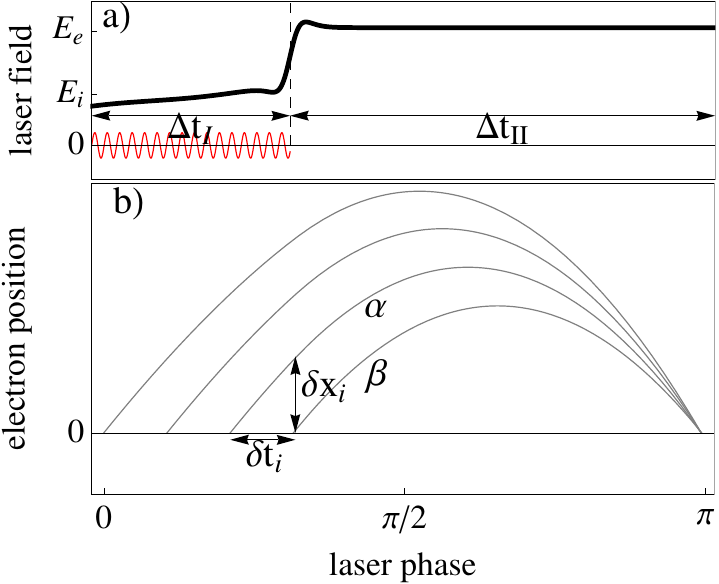}
\caption{Schematic of the recollision scenario: a) A half cycle of the tailored laser field (black). The red line is the assisting x-ray pulse. b) Different classical trajectories in the field of (a) which start into the continuum at different times but revisit the ionic core at the same time.} \label{figopti}
\end{center}
\end{figure}

It is known \cite{LEWENSTEIN1994} that the HHG process can be analyzed by taking only a few quatum orbits into account that correspond to classical trajectories. Therefore, a classical analysis is able to give first insight into our idea. In the first part of our paper, we start out by considering classical trajectories in a tailored laser field [see Fig. \ref{figopti} (a)] and find the condition when trajectories ionized at different times recollide at the same time. For the considered laser intensities, the classical trajectories have only a component along the polarization direction of the laser field, i.e. 1-dimensional trajectories are shown. We describe the principle of our method by discussing two example trajectories marked by $\alpha$ and $\beta$ in Fig. \ref{figopti} (b) being ionized separated by a small time span $\delta t_i$.  We first focus on the point in time when trajectory $\beta$ just starts. At that time, $\alpha$ has already been driven slightly away from the origin. The distance between the trajectories is $\delta x_i\approx p_i \delta t_i$ where $p_i$ is an initial momentum e.g. mediated via one-photon-ionization (atomic units are used throughout unless stated elsewise).  The momentum difference at that time is given by $\delta p_i\approx-E_i \delta t_i$ because $\alpha$ has already been decelerated by the laser field $E_i$. From now on, the momentum difference $\delta p=\delta p_i$ is conserved during the whole propagation time. The separation of both trajectories at recollision after a time $\tau$ is therefore given by $\delta x_e\approx\delta x_i+\delta p\, \tau=(p_i-E_i \tau)\delta t_i$. $\delta x_e=0$ reflects the condition on the initial momentum and on the electric field at ionization necessary for simultaneous recollision of the two trajectories: 
\begin{equation}
 E_i\approx p_i/\tau \label{recol-cond}
\end{equation}

The previous discussion applies only to the case of two single trajectories ionized with an infinitely small time separation but is sufficient for demonstration of our principle idea of simultaneous recollision for trajectories ionized from a finite time window.
%the expression is approximatly true for all trajectories leading to time-dependent $\tau$  then.
%Additionally, one can impose a second demand having an $E_i$ as large as possible to gain a large energy bandwidth of the simultaneous recolliding trajectories but on the other hand increasing $p_i$ means lowering the quantum efficiency as explained later. 

%In the following, we describe the principles how an optimal pulse shape can be found that fulfills the condition \eqref{recol-cond}. 
 In Fig. \ref{figopti}, we chose the simplest possible laser field to fulfill condition \eqref{recol-cond}: a first plateau of strength $E_i$ and duration $\Delta t_I$ which is followed by a second higher plateau of duration $\Delta t_{II}$. In this procedure, the field strength $E_e$ of the second plateau is chosen first. It determines $\tau$ for trajectories $\beta$. Then, the first plateau is found having a small slope determined by optimizing a polynomial expression to have simultaneous recollision of all classical trajectories (approximately condition \eqref{recol-cond}).  
%A schematic of one half cycle of the laser field is displayed in Fig. \ref{figopti} (a). 
%The field structure of the second plateau is does only influence the attochirp. 

Crucial for the previous scheme is the ionization with a non-zero velocity. The condition can be reached by using x rays of frequency $\omega_X$ which are responsible for ionization. In this case, the initial momentum is $p_i=\sqrt{2(\omega_x-I_p)}$ where $I_p$ is the binding potential.  The x rays co-propagate with the laser field and are spectrally filtered out before the harmonic radiation reaches the detector. $\omega_X$ can be chosen in a way (with the appropriate laser field according to Eq. \eqref{recol-cond}) to meet a resonance of a suitable absorber at $\omega_X$. Alternatively, the x rays could propagate with a tiny angle to the propagation direction of the laser.
%Therefore, we only take harmonic radiation above $\omega_X$ into account. 
%In case no appropriate filters are available, x-ray pulses could be employed which have only a noticeable intensity during $\Delta t_I$  instead of cw x-ray light.
In Fig. \ref{figopti}, we only consider trajectories ionized within $\Delta t_I$ and having a direction pointing upwards to the potential (starting in positive direction in Fig. \ref{figopti}). The initial momentum could also be directed downwards but in this case, the classical electron would not recollide and no harmonics would be emitted.
% when ionized within $\Delta t_I$ and are omitted therefore. 
A second branch of trajectories that are ignored in the figure are those trajectories emerging from the second plateau during $\Delta t_{II}$. They re-encounter the core region but all at different times and would break the desired scenario of simultaneous recollision.
%Upward trajectories emerging from the second plateau during $\Delta t_{II}$ recollide with energies below $\omega_X$ and a little fraction of the downward trajectores the ionized at the very end of it can recollide with energies in the same range as the ones above. However, these trajectories recollide at different times and would result in a significant attochirp. 
Therefore,  ionization during the second plateau has to be suppressed which demands for the use  of an x-ray pulse [sketched as red wiggeld line in Fig. \ref{figopti} (a)] rather than a cw x-ray field.  The x-ray pulse must have a non-vanishing field strength only during $\Delta t_I$ being of order of 1~fs. Moreover, as medium, we choose a gas of ions because their high ionization potential is necessary to suppress tunnel ionization. The plasma could be generated via laser ionization with a strong pre-pulse.

We briefly comment on the energy distribution of the recolliding trajectories. The classical trajectory marked by $\beta$ has a distinct role because it experiences approximately a constant laser field. Therefore, it is symmetric to the turning point and also recollides with same momentum $p_i$ as it started.  Trajectories starting prior to $\beta$ recollide  with a higher energy but at the same time as $\beta$ assuming the first plateau-like structure is chosen in agreement with Eq.~\eqref{recol-cond} as it is in Fig. \ref{figopti}. Trajectory $\beta$ plays the role of the trajectory with the lowest energy $\omega_X$.
The duration of the first plateau $\Delta t_I$ determines the velocity difference $\Delta p\approx E_i \Delta t_I$ and the energy bandwidth $\Delta \omega_q \approx \frac{1}{2}(\Delta p+p_i)^2+I_p-\omega_X=\frac{1}{2}\Delta p^2+\Delta p\, p_i$. 

%Trajectories starting after (2) are not shown because they also recollide after (2) which would result in a chirp of the attosecond pulse. Fortunately, they can be  spectrally filtered out due to their energy. Their recollision  energy  is the same as $\omega_X$ or even lower if the field changes sign as in the case a new half period starts. 
In the optimum case shown in Fig. \ref{figopti}, all trajectories recollide simultaneously. However, under real experimental conditions, deviations of the laser and x-ray field from the optimal conditions result in a non-zero time window $\Delta t_e$ of recollision. 
%In the following we discuss two reasons for non-vanishing $\Delta t_e$.
The classical model employed a discrete x-ray frequency $\omega_X$ rather than a finite bandwidth as a real pulse. The maximum allowed bandwidth can be deduced from the former model. The x rays ionize the electron with an initial velocity and arrange this way the initial displacement $\delta x_i$ between two trajectories. If $\omega_X$ deviates from its optimal value by $\delta\omega_X$, the initial momentum will deviate  by $\delta p_i\sim \delta\omega_X/(2 p_i)$ and result in a additional displacement $\delta x_i=\delta p_i \delta t_i$ which is not compensated for. Therefore, the final wave packet size is  of order $\delta p_i \Delta t_I$ resulting in a time spread of 
\begin{equation}
 \Delta t_e^{BW}=\delta p_i \Delta t_1/p=\frac{\delta\omega_X \Delta t_I}{2 p_i p} \label{limit-tBW}
\end{equation}
which has to be smaller than the envisaged pulse duration.
%Moreover, the maximum field strength of the x rays is limited in two ways: The wave packet is only allowed to be partially ionized to create HHG effectively. On the other hand the x ray field can also influence the continuum propagation if its ponderomotive potential gets too large.  Depending on the instant of ionization a classical electron can obtain a drift momentum of order $E_X/\omega_X$ causing an additional wave packet spread $\Delta x_e= E_X\tau/\omega_X $. With a final momentum of $p_e$ the spread of the wave packet in units of time is given as:
%\begin{equation}
%\Delta t_e^{E_x}= \frac{E_X\tau}{\omega_X p_e}. \label{limit-tEx}
%\end{equation}

\section{Strong-field Approximation Model}

So far, purely classical dynamics was considered.  In order to model the single-atom HHG yield, we use the strong-field approximation (SFA) and include the laser field within the dipole approximation. This way, the Fourier transformed dipole matrix element is given by \cite{FIGUEIRA2007,ufo}
\begin{equation}
 \tilde{d}_{q}=i\int_{-\infty}^\infty dt \int_{-\infty}^t dt'\int d^3\mathbf{q} \langle \Phi_0(t)\vert x\vert \mathbf{p}+\mathbf{A}(t)/c\rangle  \mathbf{E}_X(t') \langle \mathbf{p}+\mathbf{A}(t)/c\vert \mathbf{x}\vert \Phi_0(t')\rangle e^{-i S_q(\mathbf{p},t,t')} \label{M_integral}
\end{equation} 
where $\Phi_0(t)$ is the ground state wave function  of the employed zero-range potential, $S_q(\mathbf{p},t,t')=\int^t_{t'}\{[\mathbf{p}+\mathbf{A}(t'')/c]^2/2+I_p\}dt''+\omega_X t'-q\omega t$ the classical action and $q$ the harmonic number.
In the long wavelength regime the highly oscillating integral can be evaluated by using the saddle-point approximation. Expression \eqref{M_integral} is approximated by a sum 
\begin{equation}
 \tilde{d}_{q}=-i\sum_s \sqrt{\frac{(-2\pi i)^5}{\det(\tilde{S}_s)}} \langle \Phi_0(t_e)\vert x\vert \mathbf{p}_s+\mathbf{A}(t)/c\rangle  \mathbf{E}_X(t_i) \langle \mathbf{p}_s+\mathbf{A}(t)/c\vert \mathbf{x}\vert \Phi_0(t_i)\rangle e^{-i S_q(\mathbf{p}_s,t_e,t_i)} \label{M_sum}
\end{equation} 
over the saddle points $s=(\mathbf{p}_s,t_e,t_i)$ defined by
\begin{eqnarray}
\int^{t_e}_{t_i}[\mathbf{p}_{s}+A(t'')/c]dt''&=&0\\
~[\mathbf{p}_{s}+A(t_i)/c]^2/2&=&\omega_X-I_p \label{sp-ion}\\
~[\mathbf{p}_{s}+A(t_e)/c]^2/2+I_p&=&q\omega\label{sp-rec}
\end{eqnarray}
for  $\tilde{S}_{(i,j)}=\partial_i \partial_jS$ where $i,j \in \{p_x,p_y,p_z,t,t'\}$. In case $\omega_X-Ip$ is positive all saddle points are real and, therefore, $t_i$ and $t_e$ are the ionization and recollision times, respectively, being also a solution of the classical equations of motion.

%Before we investigate the temporal structure of the harmonics emitted from an ion experiencing the proposed field configuration, 
We  analyze the condition required for (near-) bandwidth-limited emission of high harmonics. The time-dependend intensity of the emitted light bursts is given by
\begin{equation}
 I(t)\propto \Big\vert\sum_j \omega_{2j+1}^2\ \vert \tilde{d}_{2j+1} \vert e^{-i\; \text{Re}\; S_{2j+1}(\mathbf{p}_s,t_e,t_i)} e^{-i\omega_{2j+1} t}\Big\vert^2
\end{equation}
where the phase $\text{Re}\; S_q(\mathbf{p}_s,t_e,t_i)\approx\alpha_0+\alpha_1 (\omega_q-\omega_c)+\frac{1}{2} \alpha_2 (\omega_q-\omega_c)^2+\ldots$ is crucial for the duration of the attosecond burst. We Taylor-expanded $\text{Re}\; S_q(\mathbf{p}_s,t_e,t_i)$ about the central harmonic frequency $\omega_c$. As outlined in \cite{MAIRESSE2003,KAZAMIAS2004} the linear coefficient of the expansion, the group delay, is simply given by 
\begin{equation}
 \alpha_1=\frac{d}{d\omega_q}\text{Re}\; S_q(\mathbf{p}_s,t_e,t_i)\bigg\vert_{\omega_q=\omega_c}\!\!\!\!\!\!=- \text{Re}\; t_e(\omega_q)\bigg\vert_{\omega_q=\omega_c}
\end{equation}
because the saddle-point equations lead to vanishing partial derivatives.
The linear chirp, the group-delay dispersion (GDD), is given by
\begin{equation}
 \alpha_2=- \text{Re}\;\frac{d}{d\omega_q} t_e(\omega_q)\bigg\vert_{\omega_q=\omega_c}\approx-\frac{\Delta t_e}{\Delta \omega_q}
\end{equation}
being in first approximation responsible for the duration of the harmonic pulse. Hence, the pulse can be considered to be bandwidth-limited if the quadratic term in the Taylor-expansion fulfills the following demand $\vert\frac{1}{2} \alpha_2 (\Delta \omega_q)^2\vert\ll 2 \pi $. This yields a criterion for the classical recollision time window $\Delta t_e$ resulting from the bandwidth of the trajectories:
\begin{equation}
 \Delta t_e\ll \frac{4 \pi}{\Delta \omega_q}=\frac{4 \pi}{\alpha}\Delta t_p \label{bw-cond}
\end{equation}
with the bandwidth-limited pulse duration $\Delta t_p=\alpha/\Delta \omega_q$  and the parameter $\alpha$ of order of unity determined by the spectrum.
This reflects that only an approximate condition $\Delta t_e\approx0$ is necessary for bandwidth-limited HHG emission.
%This way it is not necessary to realize a pulse shape that exactly results in $\Delta t_e=0$ as the one in Fig. \ref{figopti} to generate a bandwidth-limited attosecond pulse. 
In the previous section, we discussed in detail how the condition of simultaneous recollision of classical trajectories $\Delta t_e\approx0$ can be fullfilled.
We will show that it is sufficient to add only a few low-order Fourier components of the fundamental laser frequency to the laser field in order to achieve an optimized field which fullfills  Eq.~$\eqref{bw-cond}$. Moreover, Eq.~$\eqref{bw-cond}$ evidences that the chirp compensation becomes more difficult for larger bandwidth.

In our classical and quantum descriptions, we used the common assumption~\cite{xuv-ass-HHG,SCHAFER2004,TAKAHASHI2007,FIGUEIRA2007,GAARDE2005} of neglecting the x-ray field for the continuum propagation of the electron. This means dropping the term $S_x(\mathbf{p},t,t')=\int_{t'}^{t}d\tau ([p_x+A(\tau)/c]\;A_x(\tau)/c+\frac{1}{2}A_x^2(\tau)/c^2)$ in the definition of $S_q(\mathbf{p},t,t')$. We derive an approximate condition for the maximum electric field strength of the x rays being allowed that this approximation holds and that the HHG pulse duration is not influenced by the x-ray field.
When evaluating $S_x$ at the previously determined saddle points, the additional shift of the recollision times by the x-ray field is:
\begin{subequations}
\begin{align}
  t_{e,x}&=\frac{d}{d\omega_q}S_x(p_s,t_i,t_e)=\frac{\partial S_x}{\partial t_i}\frac{\partial t_i}{\partial\omega_q}+\frac{\partial S_x}{\partial t_e}\frac{\partial t_e}{\partial\omega_q}+\frac{\partial S_x}{\partial p_s}\frac{\partial p_s}{\partial\omega_q}\label{Sxa}\\
&\approx-[p+A(t_i)/c]\;A_x(t_i)/c\frac{\partial t_i}{\partial\omega_q}\approx - \frac{E_x}{\sqrt{\omega_X}}\frac{\partial t_i}{\partial\omega_q}\label{Sxb}
\end{align}
\end{subequations}
In Eq.~\eqref{Sxa} we dropped the second term because $\frac{\partial t_e}{\omega_q}$ is negligible due to the simultaneous recollision and the third term because it contains an integration over the heavily oscillating $A_x(\tau)/c$. Thus, the overall additional recollision time spread 
\begin{equation}
 \Delta t_{e,x}=\frac{E_x}{\sqrt{\omega_X}}\frac{\Delta t_I}{\Delta \omega_q}\ll\Delta t_e\label{cond-Ex}
\end{equation}
has to be much smaller than the recollision time window to justify in our case the neglection of $S_x$.

In summary, bandwidth-limited HHG emission can be reached when the recollision time window determined by the classical dynamics in the laser field fulfills Eq.~\eqref{bw-cond} and the x-ray field satisfies Eq.~\eqref{limit-tBW} and Eq.~\eqref{cond-Ex}.

\section{Generation of Attosecond Pulses}

\begin{figure}[h]
\begin{center}
\includegraphics[width=0.9\textwidth]{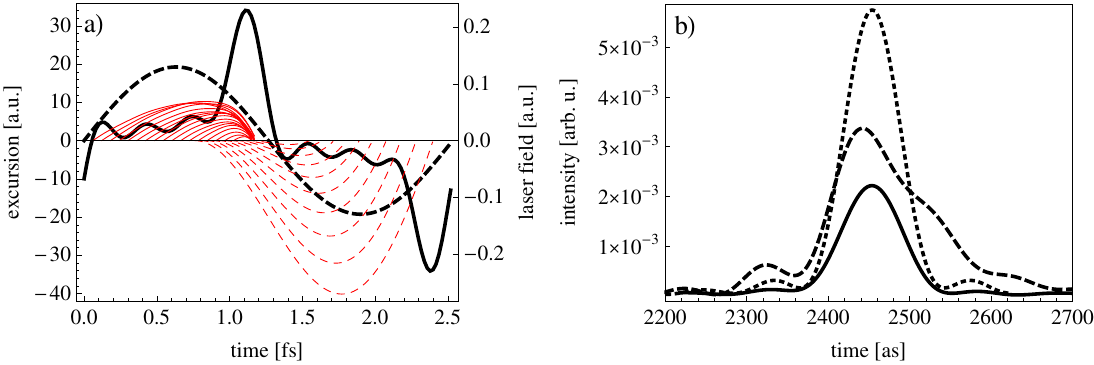}
\caption{a) Classical electron trajectories (red lines) are shown for two different field configurations (thick black lines): The dashed curves are for a conventional sinusoidal laser field whereas the full lines are for a shaped laser field consisting of $N_F=8$~harmonics that is assisted by continuous wave x-rays ($\omega_x= 60$~eV, $I_x=9\times10^{12}$~W/cm$^2$) triggering ionization.  b) Attosecond pulses emitted from the laser fields proposed in a). The solid line shows the intensity for the case of the proposed scheme whereas the dotted and dashed lines correspond to the attosecond pulses for a conventional field with a 40 eV bandwidth either with or without phase compensation, respectively.} \label{figcomparison}
\end{center}
\end{figure}

In the following, we describe an implementation of the scheme. We start from an optimal field shape determined in the way described above with a fundamental frequency of $\omega=0.06$~a.u. Then, we represent the field as a Fourier series and only take the $N_F=8$ lowest frequency components of its spectrum into account.
Following this procedure, the pulses in Fig. \ref{figcomparison} a) (solid black line) were found. Its relevant classical trajectories  ionized by a one-photon transition are shown as solid red lines. We can observe that $t_e(\omega_q)$ is approximately constant and, therefore,  we can expect bandwidth-limited harmonic emission. 
We compare it to the traditional case of a sinusoidal laser field  with similar parameters as in \cite{GOULIELMAKIS2008} (dashed lines). 
%In such a case, all energies are emitted at the same time from classical point of view promising the generation of bandwidth-limited attosecond pulses.
 Evaluating the dipole moment within the SFA  [Eq.~\eqref{M_sum}], we find the attosecond pulse generated by the fields in Fig. \ref{figcomparison} (a). The results are shown in Fig. \ref{figcomparison} (b).
The duration of the pulse generated by our method (solid line) is 86 as at full-width half maximum (FWHM) only being slightly longer than a pulse generated by a sinusoidal field in combination with a perfect chirp compensation (78 as) (dotted line). The dashed line shows the uncompensated pulse with a duration of 130 as. In all three cases, the same frequency window between 60 eV and 110 eV was employed and the x-ray field ($\omega_x= 60$~eV, $I_x=9\times10^{12}$~W/cm$^2$) was chosen to have the same average ionization rates as by tunnel ionization in the sinusoidal field. The two pulses from the sinusoidal pulse were obtained from Neon using only the short trajectories whereas in our case He$^+$ has to be employed. The laser frequency is in all cases $\omega=0.06$~a.u.  Although the single-atom emission rate of both examples are on the same order, the photon yield of our setup will be lower because the gas density is restricted by a maximum value of $5\times 10^{16}$/cm$^{-3}$ as determined later. We estimate an emitted photon number per half cycle of $10^7$ in the traditional case (density $10^{19}$/cm$^{-3}$) and $10^3$ in our case from a volume of 200~$\mu$m$~\times$~200~$\mu$m$~\times$~1~mm.

% We modify our shaped pulse to show the interplay between the pulse shape and the x-ray frequency (maybe don't keep paragraph in letter version). In Fig. \ref{fig_comp_low} we compare two shaped pulses yielding the same attosecond pulse bandwidth but at different center frequencies. The pulse denoted by the solid line emits at higher energies.
% \begin{figure}[h]
% \begin{center}
%  \includegraphics[width=0.5\textwidth]{fig_comp_low}\\
% \caption{The dasehd lines coincide with Fig. \ref{figcomparison} a) whereas the solid lines show the first half cycle of another optimized field yielding the same bandwidth but at a higher frequency. The x-ray frequency is $\omega_X=4.1$~a.u. and the same ion is employed.} \label{fig_comp_low}
% \end{center}
% \end{figure}
% The reason is that the time $t_p$ is set earlier. In order to maintain the bandwidth $E_i \Delta t_1$ has to be approximately constant. Because $\Delta t_1$ decreases, $E_i$ and, therefore, also  $p_i$ have to be increased going along with an increase of the minimum emission frequency $\omega_X$. The height of the second plateau after $t_p$ can be lowered because there is more time for recollision left when the period is kept constant. Note that this field strength of the plateau does not influence the final energy nor the bandwidth. The result shows that the spectrum can be shifted to a value that is optimal for the experimental conditions.

\begin{figure}[th]
\begin{center}
\includegraphics[width=0.9\textwidth]{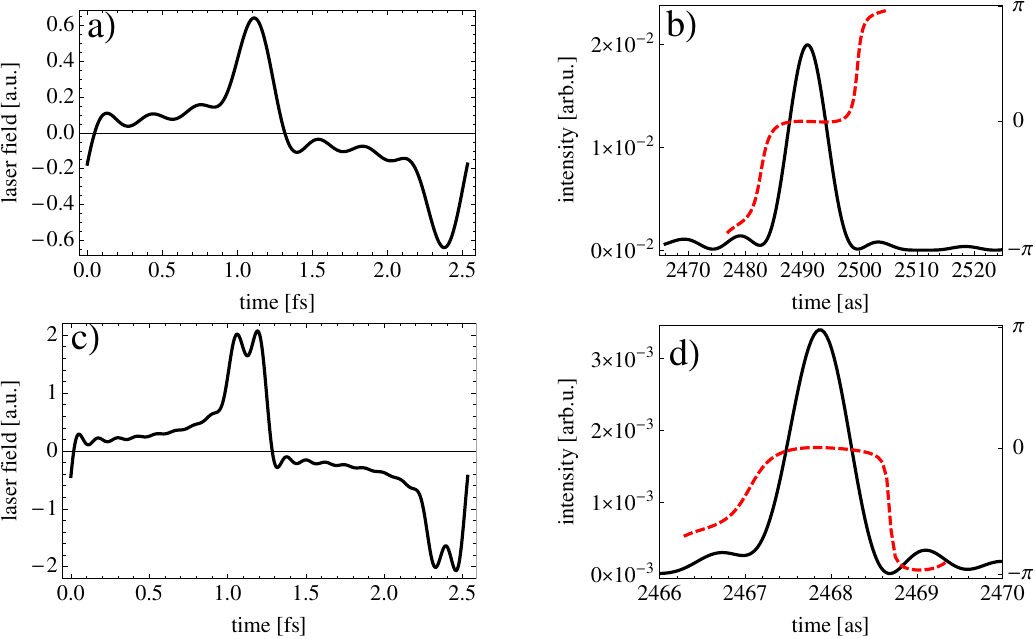}
\caption{a) shows the laser field composed of only 8 Fourier components that illuminates an Li$^{2+}$ atom with $I_p=4.5$~a.u.. The parameters of the additional x-ray field are indicated in the first row of Tab. \ref{tab_values}. The created 8~as pulse is shown in b). c) displays the laser field needed to create a pulse of 800~zs duration from Be$^{3+}$ atoms with $I_p=8$~a.u. and the parameters indicated in the second row of Tab. \ref{tab_values}. The respective pulse is shown in d). The dashed red lines are the temporal phases of the pulses.} \label{fig_highE}
\end{center}
\end{figure}

So far, the attosecond pulse production with durations little below 100~as was considered. The time difference of the uncompensated ($\sim 130$ as) and compensated ($\sim 80$ as) pulses were moderate. Now, we discuss our proposal for future experiments where the bandwidth is of order of keV and Eq.~\eqref{bw-cond} imposes a stronger constraint, demanding for a larger phase compensation over a larger frequency spectrum. 
%to shorter attosecond pulse durations being only a tiny fraction of the laser period demanding for a larger compensation over a large frequency spectrum. 
In the following, we present two examples with  production of pulses below 10~as and 1~as, see Fig.~\ref{fig_highE}.

\begin{table}[h]
\begin{tabular}{c|c|c|c|c|c|c|c}
$N_{F}$&$I_L$[W/cm$^2$]&$\omega_x$[eV]&$I_X$[W/cm$^2$]&$\Delta \omega_q$[eV]&$T_P$[as]&$\rho_{max}$[cm$^{-3}$]&$N_{ph}$\\
\hline
8&$10^{16}$&218&$3.5\times 10^{14}$&470&8&$2.5\times 10^{16}$&$10^{1}$\\
\hline
20&$10^{17}$&996&$1.4\times10^{15}$&$4.9\times10^3$ &0.8&$7\times10^{14}$&$10^{-6}$\\
\hline
\end{tabular}
\caption{Parameters for the two examples in Fig. \ref{fig_highE}: $N_F$ represents the number of Fourier components contained in the fundamental pulse, $I_L$ its peak intensity, $\omega_X$  the x-ray frequency employed for ionization, $E_X$ its field strength, $\Delta \omega_q$ the achieved HHG bandwidth, $T_P$ the HHG pulse duration, $\rho_{max}$ the estimated maximum gas density  and $N_{ph}$ an estimate of the emitted HHG photon number  per half cycle emitted from a volume of 200~$\mu$m$~\times$~200~$\mu$m$~\times$~1~mm having the maximum density.}\label{tab_values}
\end{table}
The laser field strength and the x-ray frequency have to be increased compared to the example before to obtain a much larger bandwidth and are indicated in Tab. \ref{tab_values}. This way,  attosecond pulses with a FWHM of 8~as and 800~zs are formed. The pulses are almost bandwidth-limited as can be seen from the constant phase (red dashed line) in the main part of the pulse. Here, the linear phase term given by the central frequency of the pulse was subtracted from the phase.

However, the photon yield emitted from the target is very low for several reasons: In the two examples, $\omega_X\gg I_p$ is valid because a much larger initial momentum is required, however, the ratio renders ionization inefficient.  We limited the x-ray intensity according to Eq.~\eqref{cond-Ex}.  On the other hand, one-photon ionization exhibits a larger angular distribution resulting in a large spread of the electron as soon as the initial momentum becomes larger. Finally, the gas density is limited to a small value which will be discussed shortly.

\section{Macroscopic effects}

In the remainder of this paper, we discuss consequences of applying the scheme to a macroscopic gas target. Due to dispersion, the initially optimal pulse shape will be deformed during propagation. We estimate the impact of the dispersion by investigating the pulse shape after 1-dimensional propagation through a plasma of length $L$ and refractive index $n_q=\sqrt{1-\frac{4\pi \rho_e}{\omega_q^2}}$ where $\rho_e$ is the electron density. In this case, each Fourier component of frequency $\omega_q$ propagates with the phase velocity $v_{ph}=c/n$ and an analytic expression can be obtained:
\begin{equation}
 I(t)\propto \Big\vert\sum_j \omega_{2j+1}^2\ \tilde{d}_{2j+1} e^{-i\omega_{2j+1} [t+(n_1-n_{2j+1})L/c]}\Big\vert^2
\end{equation}
The influence of the different atomic transition lines of the medium on the dispersion  is omitted because the free electron background forms the largest contribution to the dispersion. When calculating the driving pulse shape  of Fig. \ref{fig_highE} a) after a propagation length of $L=1$~mm, we find a maximum ion density of $5\times10^{16}$/cm$^3$ and $7\times10^{14}$/cm$^3$ to maintain a duration of the harmonic burst below 10~as and 1~as, respectively. Similarly, we can specify the precision of the phase of the different Fourier components that is required. The allowed fluctuations of the different components in terms of time delay is of the order of 25~as and 2.5~as, respectively, in agreement with the time delays caused by the plasma dispersion discussed before. The sensitivity is lower for harmonics with lower energies.

Apart from causing a deviation from the optimized pulse form, dispersion can also lead to phase mismatching. Due to the rather low gas density, we do not expect a dramatic phase mismatch. To achieve phase matching, we propose either to exploit the geometry of the laser focus or to use quasi-phasematching schemes as employing a weak counterpropagating IR field \cite{justin-scheme,justin-scheme2}, weak static fields~\cite{BIEGERT2010} or modulated wave guides \cite{hhg_waveguide,hhg_waveguide2,hhg_waveguide3}.

So far, the recollision scheme was discussed in the spotlight of bandwidth-limited emission of attosecond pulses. The scheme can also be applied as a new type of pump-probe technique where the atom is excited or probed at a precise time by the recolliding electron. The recollision time can be controlled via shaping the driving pulse. Moreover, the spectral diversity of the simultaneously recolliding trajectories is an excellent condition for the observation of continuum-continuum harmonics \cite{KOHLER2010}.

We thank R. Moshammer and C. Buth for fruitful discussions.

\end{document}